\documentclass[twocolumn,dvipsnames,nofootinbib,floatfix,groupedaddress,superscriptaddress]{revtex4-1}
\PassOptionsToPackage{unicode}{hyperref}
\PassOptionsToPackage{bookmarks=false}{hyperref}

\newcommand{\msw}{\text{\sc msw}}

\usepackage[utf8]{inputenc}
\usepackage{xcolor,soul}
\usepackage{amsmath}
\usepackage{amssymb,amsthm,amsfonts} 
\usepackage{bbold}
\usepackage{paralist}
\usepackage{xcolor}
\usepackage{graphicx}
\usepackage[T1]{fontenc}
\usepackage{braket} 
\usepackage[normalem]{ulem}

\usepackage{slashed} \hfuzz=2pt 
\usepackage[colorlinks=true]{hyperref}

\newcommand{\vvbar}[1]{\overset{{\scriptscriptstyle(}-{\scriptscriptstyle)}}{\nu}_{#1}}
\begin{document}

\title{Testing heavy neutral leptons produced in the supernovae explosions with future neutrino detectors}
\author{Vsevolod Syvolap}
\affiliation{Instituut-Lorentz for Theoretical Physics,
Leiden University, 2333 CA Leiden, Netherlands}

\begin{abstract}
Hypothetical particles called heavy neutral leptons (HNLs) can be produced in large quantities in the cores of supernovae during the first seconds of the explosion. These particles then decay, producing secondary energetic neutrinos that can be detected by neutrino detectors. In this paper, I identify a region of the HNL parameter space that could be tested using this method, assuming a supernova explosion at distances from 0.2 to 10 kpc. The range of HNLs  masses   $ m_N \sim 160-700$ MeV and lifetimes of $\tau_N \gtrsim 0.02$ seconds can be probed using the Hyper-Kamiokande neutrino detector. This region of the parameter space is complementary to existing bounds from primordial nucleosynthesis and to the expected sensitivity of the future SHiP experiment, thus covering  a gap in our current knowledge of HNLs up to masses of $m_N \simeq 400$ MeV.

\end{abstract}
\maketitle

\section{Introduction}

Supernova (SN) explosions provide unique environments to produce feebly interacting particles \cite{Agrawal:2021dbo} owing to large temperatures ($T \simeq \text{30 MeV}$) and high densities ($\rho \sim 10^{14} $ g/cm$^3$) in their cores \cite{Horiuchi:2018ofe}.
Specifically, supernovae explosions can produce \textit{heavy neutral leptons} (HNLs).
Phenomenologically, HNLs are heavy neutrinos whose interaction strength with intermediate vector bosons is suppressed by the flavor-dependent dimensionless constant $U_\alpha^2\ll 1$ ($\alpha = \{e,\mu,\tau\}$) \cite{Abdullahi:2022jlv}.
Production of the heavy neutral leptons during supernovae explosion has been extensively studied in the past, see Refs.~\cite{Peltoniemi1992,Raffelt:1992bs,Shi:1993ee,Kusenko:1997sp,Dolgov:2000pj,Dolgov:2000jw,Dolgov:2000ew,Abazajian:2001nj,Fuller:2003gy,Barkovich:2004jp,Hidaka:2006sg,Hidaka:2007se,Kusenko:2008gh,Fuller:2009zz,Raffelt:2011nc,Wu:2013gxa,Warren:2014qza,Zhou:2015jha,Warren:2016slz,Tamborra:2017ubu,Rembiasz:2018lok,Suliga:2019bsq,Mastrototaro:2019vug,Syvolap:2019dat}, covering the range of masses from $ \sim $ keV to hundreds of MeV) and the range of mixing angles from $U_\alpha^2 \sim \mathcal{O}(1)$ all the way to $U_\alpha^2 \sim 10^{-15}$.

Supernova constraints on the HNL parameters usually follow the energy-loss argument.
The latter is based on the expectation that the presence of additional cooling channels would shorten the duration of the SN neutrino emission.
The observations of SN1987A have shown, that neutrinos are emitted during the time interval $t \simeq 10$ sec (see Ref.\
\cite{Raffelt:2012kt} for review), which is in agreement with the existing SN explosion model \cite{Janka:2017vlw}.
At the same time, SN simulations in the presence of another feebly interacting particle -- axion -- have shown, that if the axion-driven energy emission is comparable to the rate of active neutrinos emission, the duration of neutrino signal might be reduced significantly, which would contradict the observations
\cite{Raffelt:1990yz}.
These considerations cap the energy emission in the form of any novel particle by the emission rate of one neutrino flavor,
$    \mathcal{E}_{N} \lesssim 10^{52} \text{erg/s}$.

Future neutrino detectors, such as Hyper-Kamiokande (\emph{Hyper-K}) neutrino telescope \cite{Hyper-Kamiokande:2016srs}, might provide a different way to set a constraint on HNL parameters. Unstable HNLs, produced in the SN core, could decay outside the SN neutrinosphere and produce a secondary flux of neutrinos.
Such secondary neutrinos would escape the supernova medium without interactions and be observed by the neutrino detectors. For HNLs with masses $m_N \gtrsim 200$~MeV, this secondary flux of neutrinos would have significantly higher energies, than the typical energy of SN neutrino $\braket{E_{\nu}} \simeq 10$~MeV (see e.g.\ \cite{Raffelt:2003en,Tamborra:2012ac,Janka:2017vlw}) and hence, easily distinguishable from the primary SN neutrinos signal.
This idea has been previously explored by \cite{Mastrototaro:2019vug}.

In this work, I extend the analysis of \cite{Mastrototaro:2019vug} and identify the whole region of the HNLs parameter space, where such an effect is relevant for the studies of secondary neutrinos.
I demonstrate that the Hyper-K experiment will have sensitivity to such neutrinos, able to test the mass range of hundreds of MeV.
Specifically, such a measurement would be able to close the gap between the current BBN bounds~\cite{Boyarsky:2020dzc} and the expected sensitivity region of the SHiP experiment \cite{SHiP:2018xqw} at masses, close to $m_N \gtrsim \text{ 160 MeV}$ for the HNLs coupled to the $\tau$-flavor. 
To simplify the calculations I use the snapshot-based SN model, following~\cite{Mastrototaro:2019vug,Syvolap:2019dat}.  
The snapshots are chosen as in Ref.~\cite{Mastrototaro:2019vug} to facilitate the comparison. 

The paper is organized as follows:
In the section \ref{sec:HNLs_in_SN}, a quick overview of HNL production in the SN core is described. Relevant reactions are presented and the spectra of HNL are calculated. In the section \ref{sec:HNL_decay},  decays of HNLs and spectra of secondary neutrinos are calculated for both 2- and 3-body decays. The effect of the neutrino oscillation and MSW effect is also discussed together with the final spectra of $\bar\nu_e$ at detection. In the section \ref{sec:Results}  the final result is given and comments on the different flavor mixings are provided.

\section{HNL production in supernovae}
\label{sec:HNLs_in_SN}
Heavy neutral leptons, mixed with $\nu_\tau$  can be produced in the SN media during scatterings, involving tau neutrinos in the final state. 
The relevant reactions and their matrix elements are listed in Table \ref{tab:reactions}.
\begin{table}[!t]
  \begin{center}
    \begin{tabular}{ | l | l |}
      \hline  
      Scattering process          & $S|\mathcal{M}|^2/(8G_F^2U_{\tau }^2)$                              \\ \hline
     $ \nu_\tau + \bar\nu_\tau  \to N + \bar\nu_\tau(\nu_\tau)$          & $4u(u-m_N^2)$                              \\ \hline
     $ \nu_\mu + \bar\nu_\mu  \to N + \bar\nu_\tau(\nu_\tau)$            & $u(u-m_N^2)$                              \\ \hline
     $ \nu_\tau + \nu_\tau  \to N + \nu_\tau$                  & $2s(s-m_N^2)$                              \\ \hline
     $ \bar\nu_\tau + \bar\nu_\tau  \to N + \bar\nu_\tau$      & $2s(s-m_N^2)$                              \\ \hline
     $ \nu_\tau + \nu_\mu  \to N + \nu_\mu $          & $s(s-m_N^2)$                              \\ \hline
     $ \bar\nu_\tau + \bar\nu_\mu  \to N + \bar\nu_\mu$          & $s(s-m_N^2)$                              \\ \hline
     $ \nu_\tau + \bar\nu_\mu  \to N + \bar\nu_\mu$          & $u(u-m_N^2)$                              \\ \hline
     $ \bar\nu_\tau + \nu_\mu  \to N + \nu_\mu$          & $u(u-m_N^2)$                              \\ \hline
     $ \bar\nu_\tau + \nu_e  \to N + \nu_e$          & $u(u-m_N^2)$                              \\ \hline
     $ \nu_\tau + \nu_e  \to N + \nu_e$          & $s(s-m_N^2)$                              \\ \hline
     $ \nu_e + \bar\nu_e  \to N + \bar\nu_\tau(\nu_\tau)$            & $u(u-m_N^2)$                              \\ \hline
     $ \nu_\tau + e^-  \to N + e^-$          & $4 \bar{g_L}^2s(s-m_N^2) + 4 g_R^2u(u-m_N^2) $                              \\ \hline
     $ \bar\nu_\tau + e^-  \to N + e^-$          & $4 \bar{g_L}^2u(u-m_N^2) + 4 g_R^2s(s-m_N^2) $                              \\ \hline
     $ e^- + e^+  \to N + \bar\nu_\tau(\nu_\tau)$            & $4 \bar{g_L}^2 t(t-m_N^2) + 4 g_R^2u(u-m_N^2) $                              \\ \hline
    \end{tabular}
  \end{center}
  \caption{List of reactions, that are dominant in the production of massive HNLs in the SN core. $ U^2_{\tau}$ is a mixing strength between HNL and tau-neutrino and $S$ - is a symmetry factor, $m_N$ is a mass of HNL, $g_R=\text{sin}^2\theta_W$, $\bar{g}_{L} = -1/2+\text{sin}^2\theta_W$, $\sin^2\theta_W \approx 0.22$ }
  \label{tab:reactions}
\end{table}
Note, that Refs~\cite{Fuller:2008erj,Mastrototaro:2019vug} claim that only reactions with $\nu_\mu (\bar\nu_\mu), \nu_\tau (\bar\nu_\tau)$ are relevant.
This claim is based on analysis of \cite{Buras:2002wt} where annihilation of neutrino-antineutrino pairs was studied.
Due to the large chemical potential of electron neutrinos, processes involving $\nu_e, \bar\nu_e$ are suppressed, compared to $\mu, \tau$-flavors.
However, this statement is true only if the whole neutrino spectra are taken into account.
In the current scenario, only the high-energy tail of the neutrino distribution is relevant due to the HNL mass threshold.
Therefore, the Pauli blocking is less important and the production of $\nu_e,\bar\nu_e$ annihilation is not negligible.
Apart from that, the scattering of tau neutrino/anti-neutrino on the electron neutrino and electron is important due to their larger densities and higher energies (owing to the sizeable chemical potentials).
Therefore, all these channels are included in the current analysis.\footnote{Reactions, including scatterings on electron anti-neutrinos, are indeed  suppressed due to the small number density of $\bar\nu_e$.}
Note, that mixing angle $U_{\tau}$ is considered to be small, so instead of sin$(U^2_{\tau})$, only $U^2_{\tau}$ is used. Also, matter effects could be ignored for the HNLs masses of interest \footnote{The energy of neutrino, which can undergo resonance scales as $E_{\text{res}} \sim m_N^2 $ (Ref.\cite{Notzold:1987ik}) - it is important for keV-mass HNLs as it would be of the order of temperature inside the SN core ($\sim$ MeV). But for $m_N \sim \mathcal{O} (10^2) $ MeV, this energy rises to TeV-scale and hence, matter effects can be completely neglected.}. This also means, there is no resonant production mechanism similar to MSW conversion ~\cite{Wolfenstein:1977ue,Mikheev:1986gs}. and collision production is the only option to create an HNL.
To calculate the spectrum of HNLs, {\color{black}we solve the Boltzmann equation that includes all reactions, listed in Table~\ref{tab:reactions}:}
\begin{widetext}
\begin{equation}
\label{eq:Boltzmann}
    \frac{df_N (p_N)}{dt} = \int \prod_{i=1,2,3} \frac{d^3p_i}{(2\pi)^32E_i} S|\mathcal{M}|^2 (2\pi)^4\delta(p_3 + p_N - p_1 - p_2) f_1f_2(1-f_3)(1-f_{N}).
\end{equation}
\end{widetext}
Here $f_i$ are distribution functions of in- and out-going particles, which are taken in the Fermi-Dirac form, $f_N$ -- momentum distribution of HNLs, $\mathcal{M}$ is the matrix element of the corresponding processes. 
The multiplier $1-f_N \approx 1$ since HNLs do not build up in the core. For the same reason, the inverse reaction terms are  neglected. 
Equation~\eqref{eq:Boltzmann} is solved numerically after the reduction procedure, described in~\cite{Sabti:2020yrt} which significantly simplifies the computation.

\begin{figure}[!t]
  \centering
  \includegraphics[width=\linewidth]{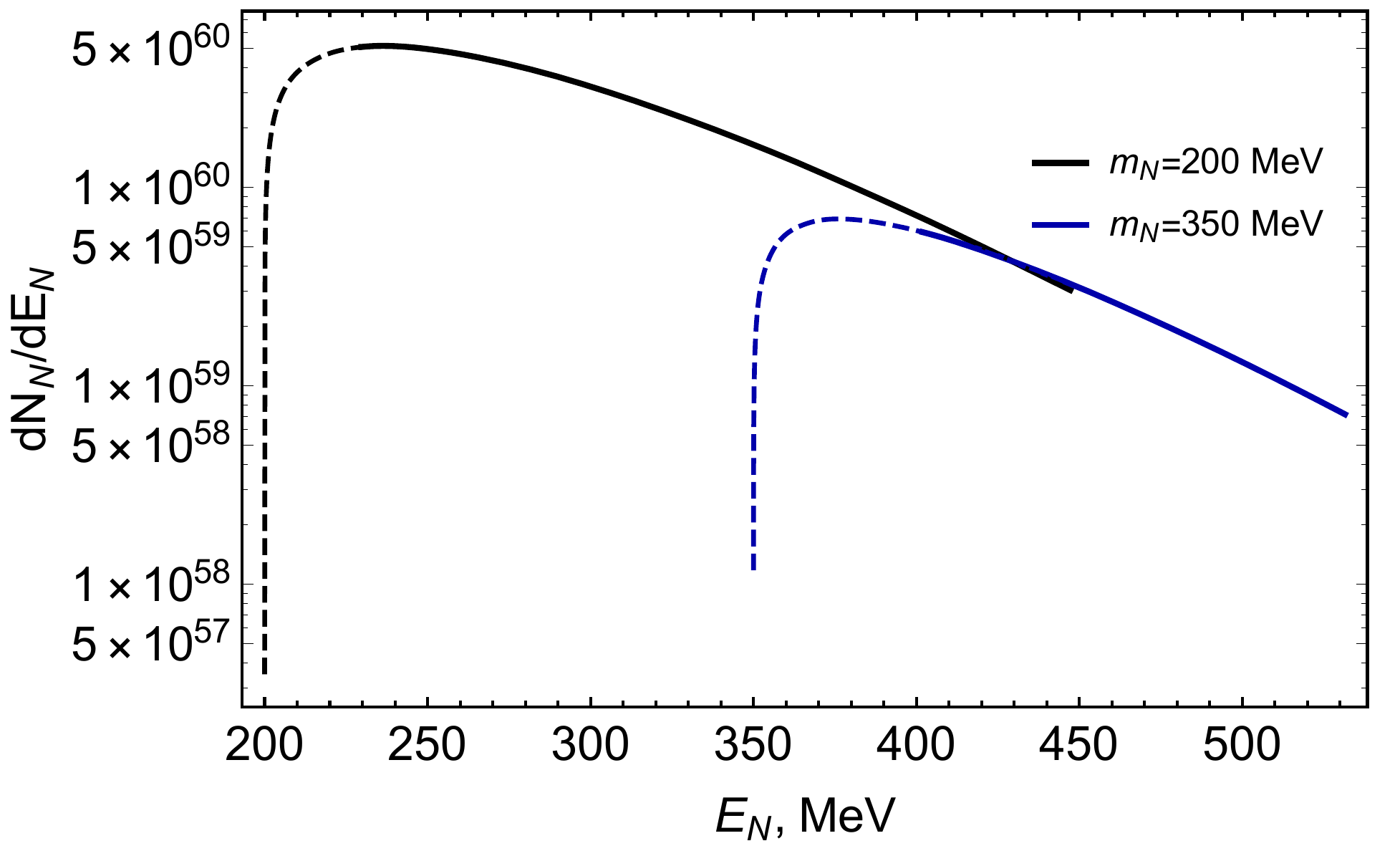}
    \label{fig:HNL_spectra}
    \caption{Energy spectra of HNLs after integration over all production regions and time interval for  HNL masses $m_N = 200, 350$ MeV in the benchmark model of the SN explosion. The mixing angle is chosen to be $U_\tau^2 = 10^{-7}$. The dashed line corresponds to the part of the HNL spectra, that is gravitationally trapped.  }
\end{figure}

There is no back-reaction, related to the change of active neutrino population since conversion occurs equally for both neutrinos and anti-neutrinos without building lepton asymmetry. For simplicity, zero chemical potentials are used for both muon and tau flavors. The chemical potential for electron neutrinos and electrons is model-dependent. For a simplified estimate, parameters $Y_{\nu_e} = 0.05, Y_{e} = 0.2$ in the core were adopted, where $Y_x$ is a relative asymmetry parameter $ Y_x = (n_x-\bar{n}_x)/n_b$ with a typical value of baryon density $n_b = 3\cdot 10^{14}$ g/cm$^3$ in the core - \cite{Fischer:2016boc,Janka:2017vlw}.
 Examples of energy spectra of HNLs are presented in Fig.\ref{fig:HNL_spectra}. 
As it is seen from spectra, most neutrinos have energy $E_N \gg m_N$. Such neutrino would be ultra-relativistic and would be able to leave the gravitational well of a proto-neutron star. But among produced HNLs, there would be a fraction, with energy smaller, than gravitational binding limit: $$E_N < E_{\text{grav}} = \frac{G M m_N}{R}$$ where $M$ - mass, enclosed within the radius $R$. Those HNLs would not be able to escape the SN core as well as their decay products. Therefore, they are ignored in calculations by cutting the energy spectra as
\begin{equation}
    \frac{d N_N}{d E_N} \to \theta(E_N-m_N-E_{\text{grav}})\frac{d N_N}{d E_N}
    \label{eq:grav_cutoff}
\end{equation}
here $\theta(x)$ - Heaviside function, $\frac{d N_N}{d E_N}$ - energy spectrum of HNLs related to previously calculated momentum spectrum as:
\begin{equation}
    \frac{d N_N}{d E_N} = \frac{E_N p_N}{2\pi^2} f_N(p_N)
\end{equation}
\section{HNLs decay}
\label{sec:HNL_decay}
The fraction of gravitationally trapped neutrinos increases with the HNL mass. After production, HNLs propagate freely outwards the SN core, where it decays. If the lifetime is long enough for HNLs to leave the neutrino-trapping region, secondary neutrinos would be also free streaming. 
Their further propagation is subject to oscillations and conversions due to the MSW effect. 
To this end, it is important to compute the spectrum of secondary neutrinos and their flavor composition.

HNLs with relevant masses and $U_\tau \neq 0$ have two kinds of dominant decay channels (see Ref.~\cite{Bondarenko:2018ptm} for review): \textit{(i)} 2-body decay  $N \to \pi^0 + \overset{\scriptscriptstyle(-)}{\nu}_{\tau}$ and \textit{(ii)} 3-body decay $N \to \vvbar\tau + \ell^- + \ell^+$, where $\ell = \{e,\mu\}$ or $N \to \vvbar\tau + \nu_\alpha + \bar\nu_\alpha$ where $\nu_\alpha$ can be any flavors.
The largest branching fraction is for the 2-body decay channel~\cite{Bondarenko:2018ptm}.
However, the 3-body decay channels can produce electron flavor (anti)neutrinos, that can be detected most efficiently (see Section~\ref{sec:Results} below).

\subsection{2-body decays}
First, consider the 2-body decay channel:
$$N \to \pi^0 + \vvbar\tau$$
In this reaction, both neutrino/anti-neutrino are available in the final state due to the Majorana nature of considered HNL. The decay width into each neutrino/anti-neutrino channel is given by:
\begin{equation}
    \Gamma_{N\to\pi+\nu} = \frac{(G_F f_\pi)^2}{32 \pi} U_{\tau}^2 m_N^3\left(1-\frac{m_\pi^2}{m_N^2}\right)^2
\end{equation}
where $f_\pi\simeq 135$ MeV is the pion decay constant. Such 2-body decay leads to a delta-function-shaped neutrino spectrum in the rest frame.
\begin{equation}
    f_\nu(E') = \frac{1}{2}\delta(E'-\overline{E}),~~~~\overline{E} = \frac{m_N^2-m_\pi^2}{2m_N}
\end{equation}
where $E'$ - the energy of the neutrino in the HNL rest frame. 
Since HNLs could be produced relativistically, even for SN temperature, smaller than the mass of HNLs, an account for the Lorenz factor is required. So in the lab frame, secondary neutrinos might have a wide range of energies. Depending on the relation between the initial HNL momentum direction and the direction to earth, the Lorenz factor will have different values. After integration over all emission angles of HNLs, it will lead to the following continuous spectra \cite{Oberauer:1993yr}:
\begin{equation}
    f_{\text{2-body}}(E) = \frac{m_N}{2\overline{E}}\text{Br}_{N \to \pi + \nu} \int_{\text{E}_{\text{min}}}^{\infty}\frac{d E_N}{p_N}\frac{dN_N}{dE_N}
      \label{eq:2-body_master}
\end{equation}
where $\frac{dN_N}{dE_N}$ energy distribution function of sterile neutrinos, $\text{Br}_{N \to \pi + \nu}$ - branching ratio for two-body decay of HNL. $E_{\text{min}}$ - minimal energy that allows for the production of an active neutrino with energy E, $$E_{\text{min}} = m_N\frac{E^2 + \overline{E}^2}{2 E \overline{E}}$$.
\subsection{3-body decays}
For 3-body decay processes the following reactions have to be considered:
\begin{equation}
    N \to \nu_a + \bar\nu_a + \vvbar\tau
    \label{eq:3-body-invis}
\end{equation}
\begin{equation}
    N \to e + e^+ + \vvbar\tau
\end{equation}
\begin{equation}
    N \to \mu + \mu^+ + \vvbar\tau
\end{equation}
Decay branching for the 3-neutrinos decay channel is:
\begin{equation}
    \Gamma_{N\to\nu_{\tau} \nu_a \bar\nu_a} = \frac{G_F^2}{768 \pi^3} U_{\tau}^2 m_N^5(1+\delta_{\tau a})
    \label{eq:3-body-nu-BR}
\end{equation}
and for decay, involving electrons:
\begin{equation}
    \Gamma_{N\to\nu_{\tau} e^+ e^-} = \frac{G_F^2}{768 \pi^3} U_{\tau}^2 m_N^5(1-4 \sin^2\theta_\text{W}+8 \sin^4\theta_\text{W})
    \label{eq:3-body-e-BR}
\end{equation}
In the case of $\mu^\pm$ decay products, the branching ratio is  suppressed, compared to \ref{eq:3-body-e-BR} for smaller masses, and becomes the same, for $m_N \gg m_\mu$ \cite{Bondarenko:2018ptm}. In either case, only tau-neutrinos can be produced in this channel while its branching is suppressed, compared to 2-body decays. 
Therefore,  decays into either $e^-e^+$ or $\mu^-\mu^+$ pairs might be ignored.
The quoted decay widths are the same for $\nu_\tau$ and $\bar\nu_\tau$, owing to the Majorana nature of HNL. 
 In the rest frame, the double-differential (for angle and energy) spectrum is of one of two types -- \cite{Mastrototaro:2019vug}
\begin{equation}
    \frac{d^2N}{dEd\Omega} \equiv f_I(E)=\frac{256}{54}\frac{E^2}{m_N^3}\left(3-4\frac{E}{m_N}\right)
\end{equation}
\begin{equation}
    f_{II}(E)=48\frac{E^2}{m_N^3}\left(1-2\frac{E}{m_N}\right)
\end{equation}
spectrum $f_I$ corresponds for both of neutrinos in the  process $N \to \nu_a + \bar\nu_a + \nu_\tau$ while the spectrum $f_{II}$ corresponds for anti-neutrino in it. In the case of charged conjugated processes, functions $f_I$ and $f_{II}$ have to be swapped. $E$ is the energy in the center of the mass frame.
A similar procedure of integrating over all angles and HNL energies must be repeated here, giving the  final energy spectrum:
\begin{widetext}
\begin{equation}
    f_{\text{3-body}}(E) = \int dE_N\int d \cos\theta \,\text{Br}_{3\nu, a} \frac{dN_N}{dE_N} f_{a}\left(\frac{E}{\gamma(1+\beta \cos\theta)}\right)
    \label{eq:3-body_master}
\end{equation}
\end{widetext}
where $\text{Br}_{3\nu, a}$ - branching ratio of the \ref{eq:3-body-invis} process, $\gamma, \beta$ - Lorenz factor and velocity to speed of light ratio and $f_{a}$ - one of $f_I, f_{II}$ depending on the particle and process. The argument of $f_{a}$ comes from Lorenz's boost of HNL in the correspondent direction.

\subsection{Secondary neutrino spectrum on Earth}
The Hyper-K detection  is most efficient for detecting electron anti-neutrinos, \cite{Sekiya:2017lgj,Wilson:2021uwb}. 
Therefore, we need to compute the composition of the secondary $\bar\nu_e$ neutrinos on Earth.
Most of the secondary neutrinos from the HNL decays have tau-flavor from two-body decays. 
However, while propagating outwards the SN media, they might undergo either flavor conversions via MSW effect \cite{Mikheev:1986gs,Wolfenstein:1977ue} and more complicated collective effects that were studied, for example, in Ref.\cite{Airen:2018nvp,Duan:2010bg}, or vacuum neutrino oscillations, see Ref.~\cite{Giganti:2017fhf}. 
This  leads to significant differences in the neutrino flavor composition at detection and on Earth.

The exact mechanism of flavor conversion depends on the position of HNL decay and SN density profile. 
It can be roughly separated into three regions, depending on what distance from the SN center,  $r_{\text{dec}}$, the decay took place, \cite{Mirizzi:2015eza}.
\begin{itemize}
    \item[Region I:] Collective neutrino oscillations area: closer to the core of the SN, $r_{\text{dec}} \simeq \mathcal{O} (100) $~km.
    Ref.~\cite{Lunardini:2012ne} demonstrated that the contribution of the collective effects to the final $\vvbar{e}$ spectrum is negligible, compared to the MSW conversion (see below).
    Therefore I will ignore the change in the flavor composition in Region~I.

    \item[Region II:] MSW conversions area extends out to the radii $R_{\msw} \sim 10^5 \text{ km}$. For HNLs that decayed between $\mathcal{O} (\text{ 100 km}) \lesssim r_{\text{dec}} \lesssim R_{\msw}$ their decay products will experience MSW conversion. 
    Notice that secondary neutrinos produced in Region~I would also experience MSW conversions.
    \item[Region III:] Vacuum oscillations area: for larger radii, $ r_{\text{dec}} \gtrsim R_{\msw}$, matter effects become insignificant and neutrino flavour oscillations are as in vacuum. 
    Those neutrinos that undergo MSW conversions in Region~II do not oscillate, since they are produced in a given mass eigenstate.
\end{itemize}
HNLs are produced in the SN core whose size is ($R \approx \text{ 10 km}$) is much smaller than  
the other relevant  scales. 
Therefore, the evaluation of the final spectrum of secondary neutrinos depends on the ratio $v\tau_N \gamma/R_{\msw}$ (where $v$ is the HNL's velocity, $\tau_N$ is the lifetime and $\gamma$ -- the Lorentz gamma-factor).
For a given mass and the mixing angle of HNL, this ratio depends only on the energy of the HNL.
Let us introduce the energy of HNL for which $v\tau_N \gamma = R_{\msw}$:
\begin{equation}
\label{eq:epsilon_N}
    \frac{\epsilon_N}{m_N}\frac{\sqrt{\epsilon_N^2-m_N^2}}{m_N}\tau_N = R_{\msw}
\end{equation}
Note, that for the lifetimes of interest (constrained from BBN to be $\tau_N \lesssim 0.02 - 0.05$~sec, \cite{Boyarsky:2020dzc}), the traveled distance of HNLs remains smaller than  $R_{\msw}$.
Indeed, $c \tau_N  = 1.2 \times 10^4 \text{ km}$) for $\tau_N = \text{0.05 sec}$. 
Hence, most of the HNLs decay deeply in Region~II.

HNLs with energy $E_N \lesssim \epsilon_N$ decay in the MSW area (Region~II).\footnote{As mentioned above we neglect composition change due to collective neutrino oscillations in the Region~I} 
The resulting $\bar\nu_e$ spectrum  is given by (c.f.~\cite{Mirizzi:2015eza}):
\begin{equation}
\label{eq:RegII}
    F_{\bar\nu_e, \text{MSW}} = \bar P_{ee}F_{\bar\nu_e, \text{II}}^0 + [1-\bar P_{ee}]F_{\bar\nu_x, \mathrm{II}}^0
\end{equation}
where $F_{\bar\nu_e, \text{II}}^0$ --  flux of the secondary $\bar\nu_e$ in the Region~II,  $F_{\bar\nu_x, \text{MSW}}^0$ -- flux of  muon or tau-flavor secondary neutrinos. 
Survival probability $\bar P_{ee}$ depends on the hierarchy of neutrino masses: $\bar P_{ee} = \cos^2\theta_{12} \approx 0.692$ for NH and $\bar P_{ee} = 0$ for IH. 

\begin{figure}
    \centering
    \includegraphics[width=0.5 \textwidth]{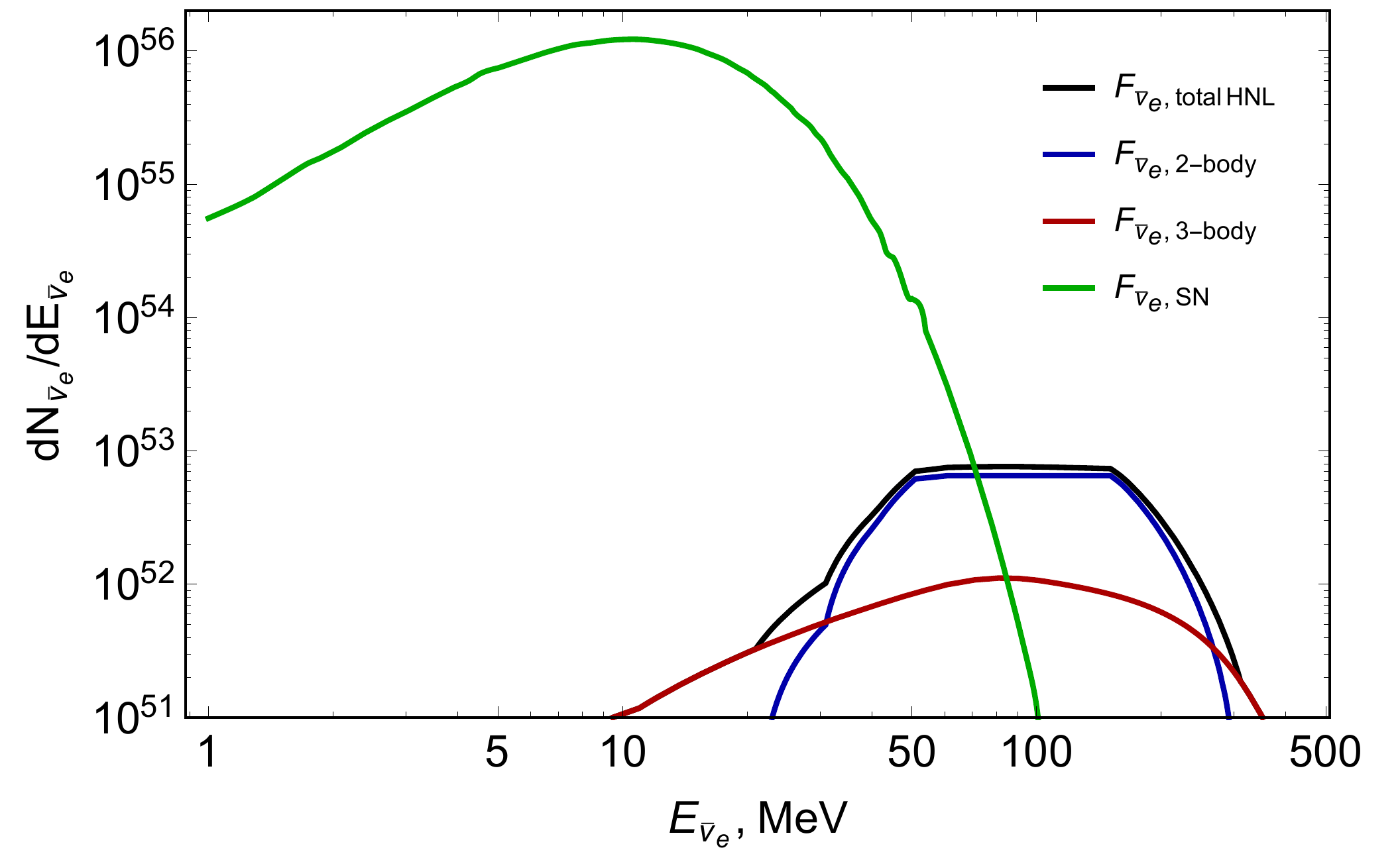} 
     \includegraphics[width=0.5 \textwidth]{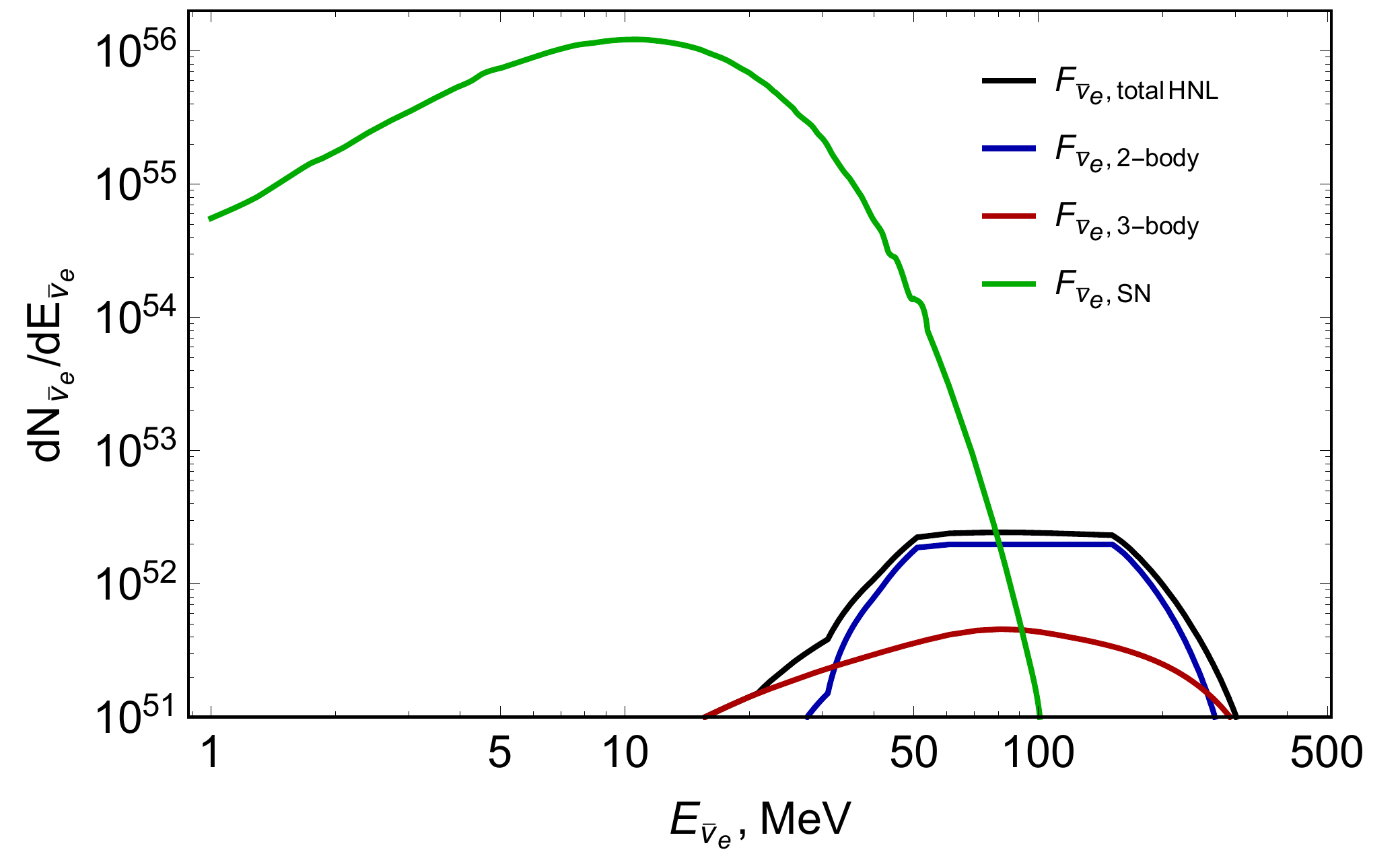}
    \caption{Spectra of $\bar\nu_e$ at Earth, after flavor conversions in two cases of inverted \textit{(Top plot) } and normal \textit{(Bottom plot)} hierarchies. Here we used $m_N = 300$ MeV and $U_{\tau}^2 = 10^{-7}$. Neutrinos, originating from 2- and 3- body decays are separated to demonstrate a distinct feature of the 2-body decay spectra. For a range of energies, spectra of neutrinos appear to be flat due to cut-off \ref{eq:grav_cutoff}. In the case of detection, such a feature can be a signal for gravitationally trapped heavy novel particles. Estimated spectra of SN electron antineutrinos are presented for comparison. It is seen, that although the mean energies of both spectra differ significantly, only at energy $E_{\nu_e} \sim 80-90$ MeV, the secondary neutrino spectra becomes dominant. }
    \label{fig:nue_specra}
\end{figure}

HNLs with energies $E_N\gtrsim\epsilon_N$ would predominantly decay in Region~III. 
The resulting $\bar\nu_e$ spectra are given by:
\begin{equation}
\label{eq:RegIII}
    F_{\bar\nu_e, \text{osc}} = P^{\text{osc}}_{ee} F_{\bar\nu_e, \text{III}}^0 + P^{\text{osc}}_{e\mu} F_{\bar\nu_\mu, \text{III}}^0 + P^{\text{osc}}_{e \tau} F_{\bar\nu_\tau, \text{III}}^0
\end{equation}
where $P^{\text{osc}}_{ee}$ -- survival probability for electron neutrino, $P^{\text{osc}}_{e(\mu/\tau)}$ - oscillation probability from a corresponding flavor, $F_{\bar\nu_\alpha, \text{III}}^0$ for $\alpha = \{e,\mu,\tau\}$ - spectra of the secondary neutrinos, produced from this higher-energy part of the HNL spectra. Note, that the subscript ``osc''  underlines their production in the vacuum oscillation area. The standard expression for vacuum oscillating, time-averaged probabilities, for the case of vanishing complex phases in the PMNS matrix (see Ref.\cite{Mirizzi:2015eza}), is:
\begin{equation}
    P^{\text{osc}}_{\nu_\alpha \to \nu_\beta} = \delta_{\alpha\beta} - 2 \sum_{i>j} U_{\alpha i}U_{\alpha j} U_{\beta i} U_{\beta j}
\end{equation} 
here $U_{\alpha i}$ are the components of PMNS matrix (see Ref.\cite{Giganti:2017fhf}).
For the PMNS mixing angles, I take their best-fit values from \cite{Capozzi:2013csa}:
\begin{eqnarray}
\sin^2\theta_{12} = 0.308 \nonumber\\
\sin^2\theta_{13} = 0.023\nonumber\\
\sin^2\theta_{23} = 0.437\nonumber
\end{eqnarray}

Spectra~\eqref{eq:RegII} and \eqref{eq:RegIII} are combined to obtain the final $\bar\nu_{e}$ spectrum.
Examples of resulting electron anti-neutrino spectra for both normal and inverted hierarchies are presented in Fig.~\ref{fig:nue_specra}.

\section{Main results}
 \label{sec:Results}
\begin{figure}[!t]
  \centering
  \includegraphics[width=\linewidth]{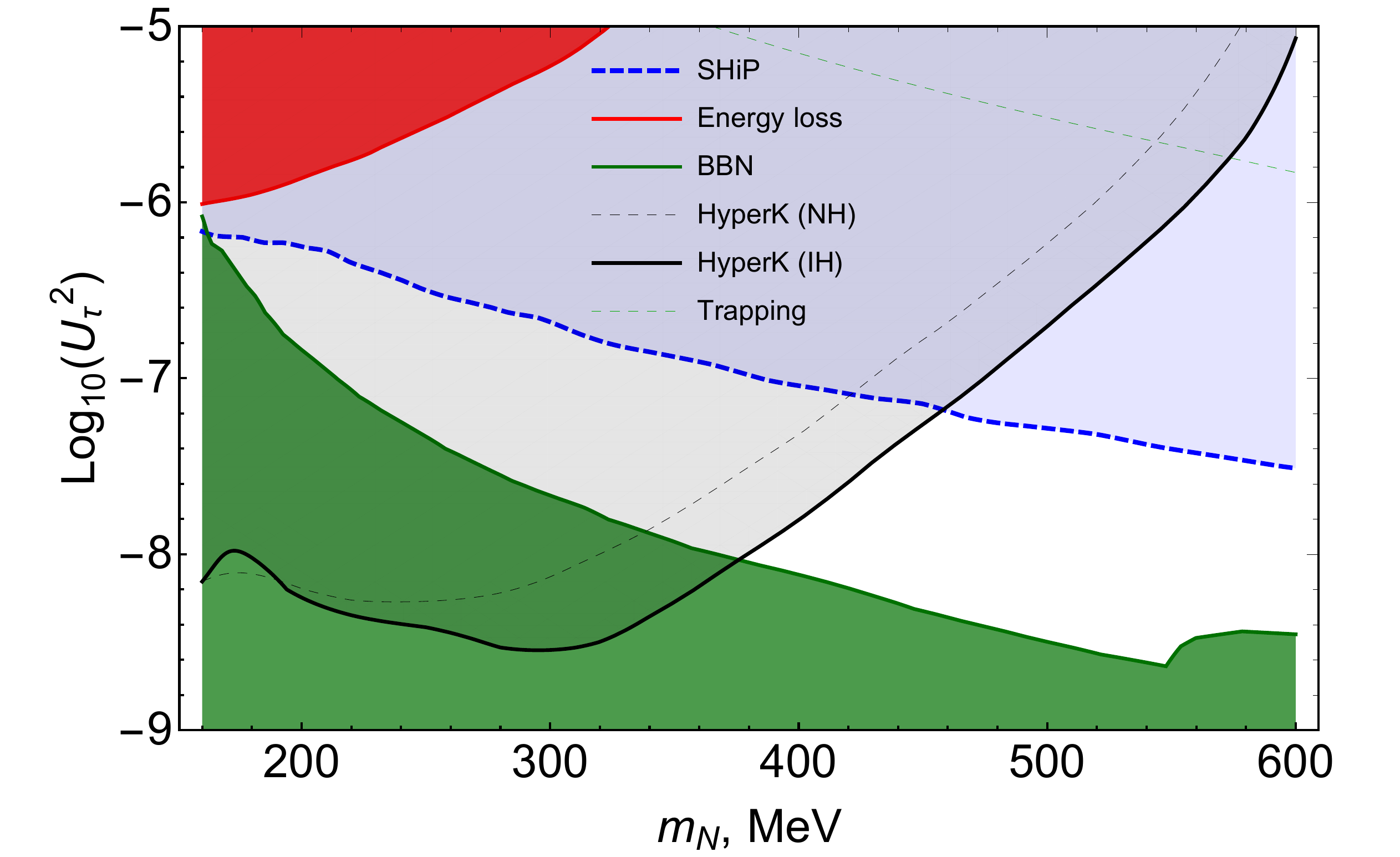}
    \caption{\textbf{Main result:} a region of the HNL parameters space, testable by the Hyper-K detector.
    The shaded grey area (the main result) shows the region in which more than 100 detection events are expected for the detector tank with a capacity of 0.22~Mton. 
    The energy-loss bound \cite{Mastrototaro:2019vug} (red region) is presented for comparison.
     The green dashed line corresponds to HNLs with a decay path < 10 km, which would decay before leaving the core and/or the neutrinosphere, leaving the decay products trapped. 
     For a wide range of masses, the gap between existing BBN bounds~\cite{Boyarsky:2020dzc} and  the potential reach of the SHiP experiment~\cite{SHiP:2018xqw} will be closed.  
     \label{fig:results}}
\end{figure}

Using the approach, described in Sections~\ref{sec:HNLs_in_SN} and \ref{sec:HNL_decay}, fluxes of secondary active neutrinos from HNLs decays were obtained. 
Taking into account possible flavor conversions, those neutrinos were traced to Earth, where the resulting electron anti-neutrino flux was calculated. 
To estimate the detection rate, the inverse beta decay (IBD) process is considered, which is the most effective in Hyper-K for neutrino detection:
\begin{equation}
    \bar\nu_e+p\to n+e^+
\end{equation}
The number of events with this process can be estimated as:
\begin{equation}
\label{eq:Nevents}
    N_{\text{events}} = \int_{E_c} dE_{\bar\nu_e} \sigma_{\text{\sc ibd}}(E_{\bar\nu_e}) \frac{dN_{\bar\nu_e}/dE_{\bar\nu_e}}{4\pi R_{\mathrm{SN}}^2} N_c
\end{equation}
here $\sigma_{\text{\sc ibd}}(E_{\bar\nu_e})$ is the cross-section for inverse beta-decay \cite{Strumia:2003zx}, $R_{\text{SN}}$ - distance to the SN. For the estimate, distance is taken  $R_{\text{SN}} = \text{ 10 kpc}$ (which roughly corresponds to the distance to the Galactic Center). $N_c$ -- the number of interacting centers (hydrogen atoms). Such an estimate is in good agreement with the number of events, presented in \cite{Wilson:2021uwb} for a supernova neutrino flux. 
To separate the contribution of the original SN neutrinos, the integration in \eqref{eq:Nevents} starts from $E_{c} \approx \text{80 MeV}$.
At this energy, the spectrum of secondary neutrinos becomes dominant, while the original $\bar\nu_e$ spectra exponentially drop, as Figs.~\ref{fig:nue_specra} demonstrate. 
The main result is presented in Fig.~\ref{fig:results}.
It shows that part of the HNL parameter space, unexplored by the current and future experiments, can be  tested by the Hyper-K detector, should a supernova explosion occur within the radius of $\lesssim \text{ 10 kpc}$. 
The black lines in Fig.~\ref{fig:results} are drawn for $N_{\text{events}}= 100$.
 \bigskip

\section{Conclusion and discussion}
An explosion of a supernova at distances $\lesssim \text{ 10 kpc}$ would lead to multiple events detected by the  Hyper-Kamiokande detector. 
Alongside neutrino physics, it would provide a test of a significantly larger region in the parameter space of heavy neutral leptons, than the one, given by current energy-loss constraints. For masses between $m_N \approx 160$ MeV and $m_N\approx 450$ MeV, it would close a gap between the sensitivity of the future SHiP experiment~\cite{SHiP:2018xqw} and current BBN bounds~\cite{Boyarsky:2020dzc}, completely covering the parameter space of HNLs at these masses. 
These constraints are much more direct than energy-loss arguments and the expected future improvement of the neutrino detectors would make them even more competitive (should even a single supernova explode sufficiently nearby).  
\begin{figure*}[h!]
  \centering
  \includegraphics[width=0.5\linewidth]{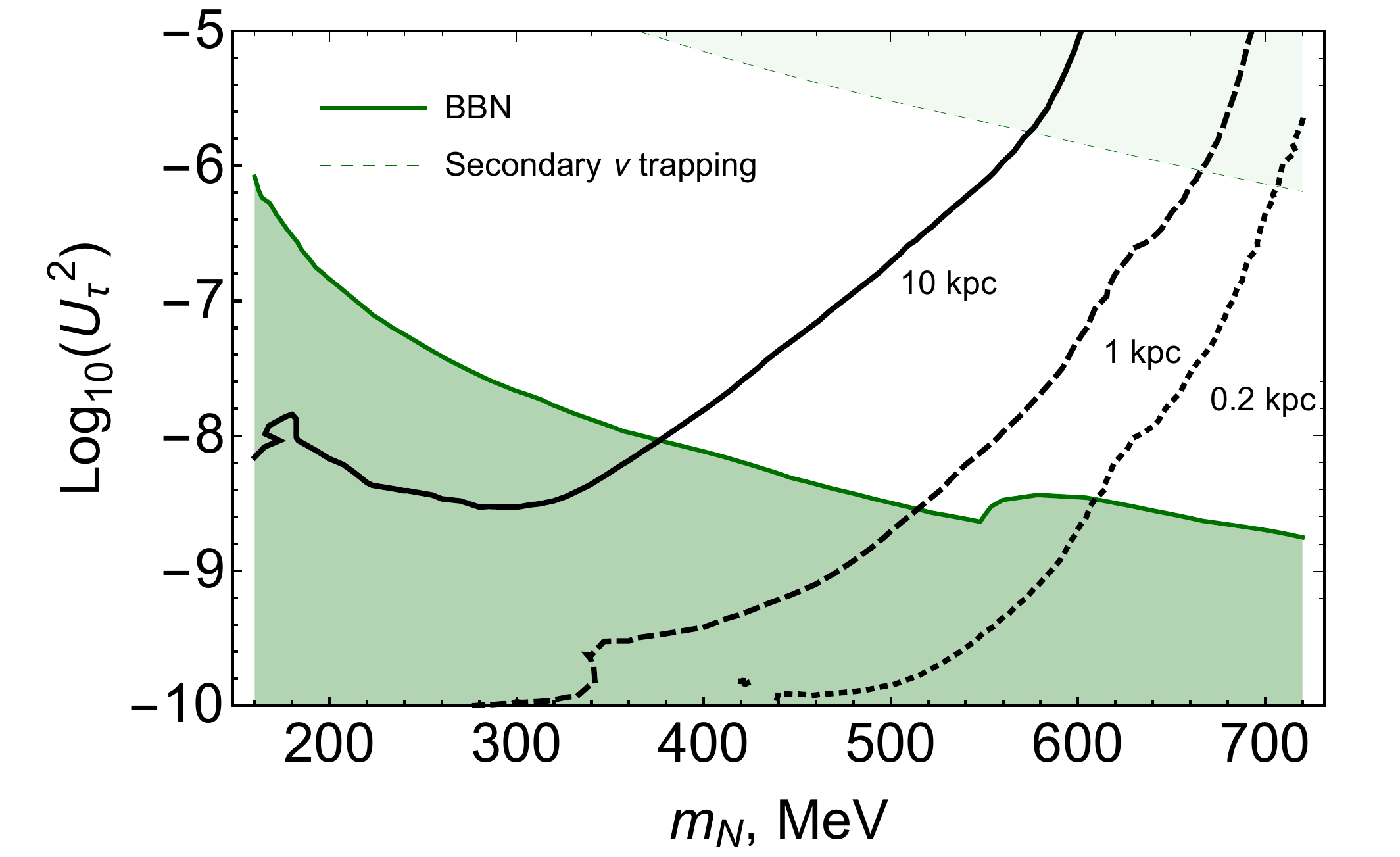}~\includegraphics[width=0.5\linewidth]{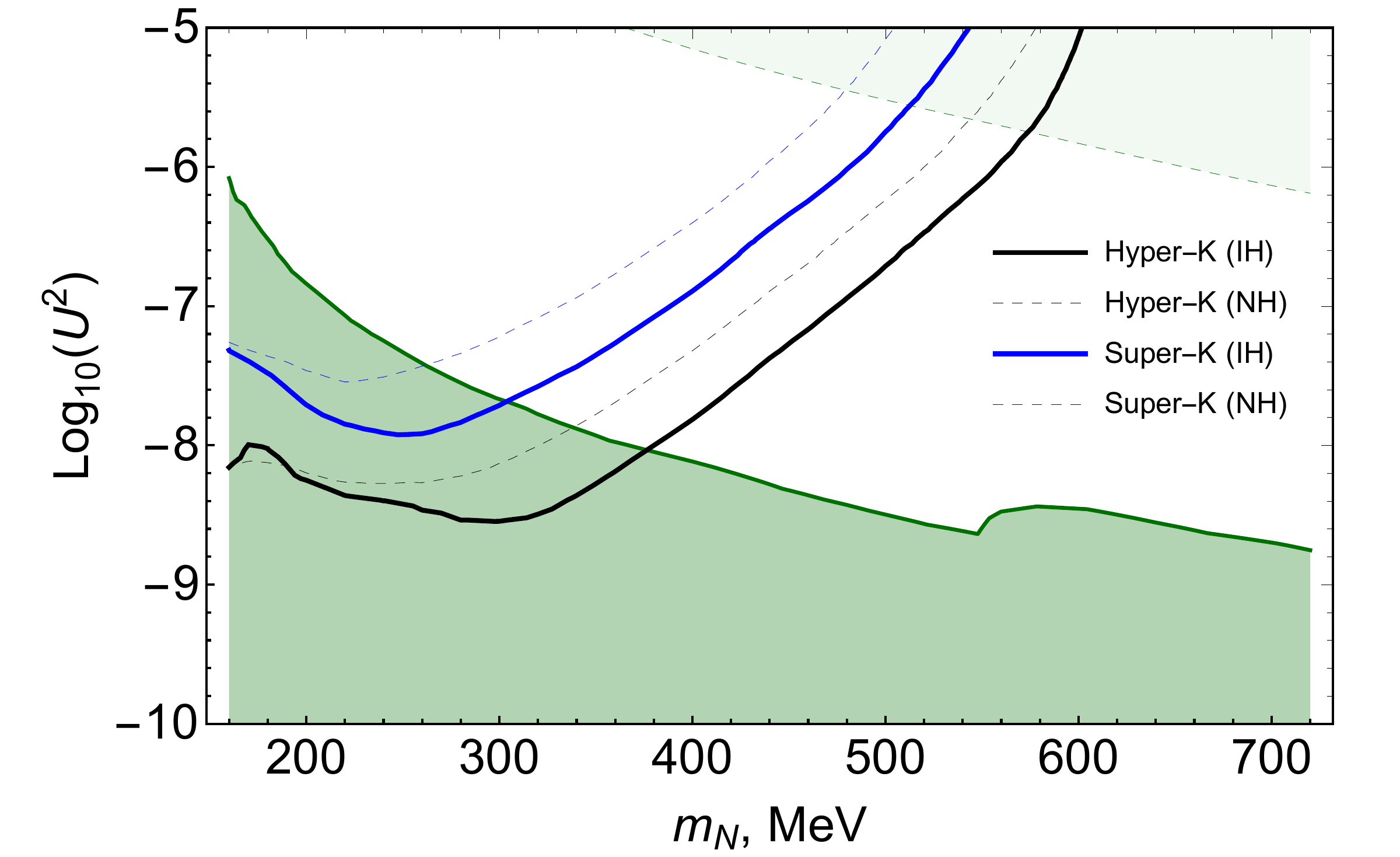}
    \caption{\textbf{Left plot:} Contours of $N_{events} = 100$ detection events corresponding for different distances to a supernova - 10 (Galactic Center), 1 and 0.2 kpc (Betelgeuse). An inverted hierarchy is assumed. Contours do not reach  higher masses even for the nearest SN, since HNLs are not capable of escaping the core. \textbf{Right plot:} Contours of same detection events for the case of Super-Kamiokande detector of both neutrino hierarchies compared to the Hyper-K case at a distance of 10 kpc.
    }
    \label{fig:several_distances}
\end{figure*}

This result, of course,  strongly depends on the distance to the SN,  Fig.~\ref{fig:several_distances}. 
For nearby explosions (for example, that of Betelgeuse located $d\approx 200$~pc) would lead to several orders of magnitude more events and can be detected even with the existing Super-Kamiokande (Super-K) detector, see Fig.~\ref{fig:several_distances}, right.
Due to the smaller detector capacity, the number of supernova neutrinos in Super-K (see Ref. \cite{Suwa:2019svl})  is estimated to be at the level of a few $\times 10^3$   for an SN in the Galactic Center. 
This corresponds to roughly an order of magnitude smaller sensitivity than Hyper-K. 
Therefore, our result for Hyper-K should be weakened by an order of magnitude in terms of $U^2$ to obtain a similar estimate for Super-K -- see the right plot in Fig.~\ref{fig:several_distances}.

Can observations of SN1987A provide valuable limits on the HNL parameters? 
Kamiokande-II detected only 12 events, while Hyper-K would detect $\mathcal{O}(10^5)$ events for the SN at this distance in the energy range 10-40 MeV \cite{PhysRevLett.58.1490}.
Therefore the sensitivity for Kamiokande-II was about 4 orders of magnitude weaker than that of Hyper-K. 
The sizeable number of secondary neutrinos could only be produced for the mixing angles that were 4 orders of magnitude \textit{larger}.
Such HNLs would be short-lived and decay inside the core or neutrinosphere.


It is also possible to consider the scenario, where HNLs are mixed not only with tau flavor.
In the case of \textit{$\mu$-flavor mixing}, it can be expected that the production rate of HNLs would be the same, if we substitute $U_\tau^2 \to(U_{\mu}^2+ U_{\tau}^2)$. 
Decays of such HNLs would also proceed faster due to more available decay channels (including muons and charged pions above the corresponding mass thresholds). 
However, the secondary neutrino flux would remain largely the same. 
Hence, the resulting region will be mostly provided by the larger mixing angles $(U_{\mu}^2,U_{\tau}^2)$ similar to the one, obtained in Sec.~\ref{sec:Results}. 

In the case of \textit{$e$-flavor mixing} the production is more complicated since electron flavor is degenerate in the SN core. There are significantly more electron neutrinos than anti-neutrinos. Therefore, they will be produced in  $\nu_e$'s interactions and will depend strongly on their number density and hence, on the electron-asymmetry radial profile in the SN. However, larger chemical potential and hence, larger electron neutrino energies could allow for the production of heavier HNLs. I leave this study for future work as it requires more model-dependent parameters taken into account.
Additionally, electron flavor mixing allows for the production of secondary electron flavor neutrinos in 2-body decays, which does not admit the simple treatment as in the case of $\mu$ flavor mixing. 
Finally, for  both $e$- and $\mu$-mixings there is no gap between SHiP region and BBN constraints \cite{Boyarsky:2020dzc} until large masses of around $m_N \approx 700$ MeV~\cite{Bondarenko:2021cpc}. 
Hence secondary SN neutrino will not explore regions of the HNL parameter space.

Finally, it should be mentioned, that this result is strictly valid \textit{only} if the SN remnant is  a neutron star.  The emission profile of neutrinos as well as parameters of the SN environment in the case of the black hole formation could be different and requires a separate study.

\section{Acknowledgments}
I would like to thank Dr.~Oleg Ruchayskiy and Dr.~Subodh Patil for their help in writing this paper, the productive discussions, and the comments I received to improve this work.

\bibliography{preamble,sample}

\end{document}